\documentclass{emulateapj}

\countdef\decade=200
\advance\decade by \year
\countdef\hours=201
\advance\hours by \time
\divide\hours by 60
\countdef\mins=202
\advance\mins by \hours
\multiply\mins by 60
\multiply\hours by 100
\countdef\miltime=203
\advance\miltime by \hours
\advance\miltime by \time
\advance\miltime by -\mins

\newcommand{\ltsim}{\protect\raisebox{-0.5ex}{$\:\stackrel{\textstyle <}
        {\sim}\:$}}
\newcommand{\gtsim}{\protect\raisebox{-0.5ex}{$\:\stackrel{\textstyle >}
        {\sim}\:$}}

\newcommand{\msun}{M_{\odot}}
\newcommand{\lsun}{L_{\odot}}
\newcommand{\sfr}{\dot{M}_*}

\newcommand{\mx}{M_{\rm X}}
\newcommand{\mirdc}{M_{\rm IRDC}}
\newcommand{\mhcn}{M_{\rm HCN}}
\newcommand{\mcs}{M_{\rm CS}}

\newcommand{\fx}{f_{\rm X}}
\newcommand{\firdc}{f_{\rm IRDC}}
\newcommand{\fcs}{f_{\rm CS}}
\newcommand{\fhcn}{f_{\rm HCN}}

\newcommand{\tff}{t_{\rm ff}}
\newcommand{\tffx}{t_{\rm ff-X}}
\newcommand{\tffcs}{t_{\rm ff-CS}}
\newcommand{\tffhcn}{t_{\rm ff-HCN}}
\newcommand{\tffirdc}{t_{\rm ff-IRDC}}

\newcommand{\sfrff}{\mbox{SFR}_{\rm ff}}
\newcommand{\sfrffx}{\mbox{SFR}_{\rm ff-X}}
\newcommand{\sfrffhcn}{\mbox{SFR}_{\rm ff-HCN}}
\newcommand{\sfrffirdc}{\mbox{SFR}_{\rm ff-IRDC}}
\newcommand{\sfrffcs}{\mbox{SFR}_{\rm ff-CS}}

\newcommand{\avir}{\alpha_{\rm vir}}
\newcommand{\calm}{\mathcal{M}}
\newcommand{\tad}{t_{\rm AD}}
\newcommand{\ecore}{\epsilon_{\rm c}}

\slugcomment{Accepted to the Astrophysical Journal, September 12, 2006}

\begin{document}

\title{Slow Star Formation in Dense Gas: Evidence and Implications}

%\centerline{DRAFT: \today}

\author{Mark R. Krumholz\footnote{Hubble Fellow}}
\affil{Department of Astrophysical Sciences, Princeton University,
Princeton, NJ 08544}

\author{Jonathan C. Tan}
\affil{Department of Astronomy, University of Florida, Gainesville, FL
32611}

\begin{abstract}
It has been known for more than 30 years that star formation in giant
molecular clouds (GMCs) is slow, in the sense that only $\sim 1\%$ of
the gas forms stars every free-fall time. This result is entirely
independent of any particular model of molecular cloud lifetime or
evolution. Here we survey observational data on
higher density objects in the interstellar medium, including infrared
dark clouds and dense molecular clumps, to determine if these objects
form stars slowly like GMCs, or rapidly, converting a significant
fraction of their mass into stars in one free-fall time. We find no
evidence for a transition from slow to rapid star formation in
structures covering three orders of magnitude in density. This has
important implications for models of star formation, since competing
models make differing predictions for the characteristic density at
which star formation should transition from slow to rapid. The data
are inconsistent with models that predict that star clusters form
rapidly and in free-fall collapse. Magnetic- and turbulence-regulated
star formation models can reproduce the observations qualitatively,
and the turbulence-regulated star formation model of Krumholz \&
McKee quantitatively reproduces the infrared-HCN luminosity correlation
recently reported by Gao \& Solomon. Slow star formation also implies
that the process of star cluster formation cannot be one of global
collapse, but must instead proceed over many free-fall times. This
suggests that turbulence in star-forming clumps must be driven, and
that the competitive accretion mechanism does not operate in typical
cluster-forming molecular clumps.
\end{abstract}


\section{Introduction}

\citet{zuckerman74} first pointed out that star formation in giant
molecular clouds (GMCs) happens surprisingly slowly. Comparing the mass of
GMCs in the Galaxy with the total Galactic star formation rate implies
that no more than $\sim 1\%$ of the gas can form stars for each
cloud free-fall time. This result is sufficiently surprising that
numerous theories have been proposed to explain it, ranging from the
idea that strong magnetic fields \citep[e.g.][]{allen00} or turbulence
\citep[e.g.][]{krumholz05c} within clouds inhibit star formation to
the idea that galactic-scale gravitational instability regulates star
formation \citep[e.g.][]{li05a} to the idea that GMCs are, contrary to
most observational estimates to date \citep{blitz06a}, actually
gravitationally unbound \citep[e.g.][]{clark04}.

Following \citet{krumholz05c}, we define the dimensionless star
formation rate per free-fall time $\sfrff$ as the fraction of an
object's mass that it converts into stars per free-fall time at the
mean density of that object. The \citet{zuckerman74} argument shows
that $\sfrff\approx 0.01$ for GMCs. This provides a powerful
constraint on models of GMCs. For example, it rules out the early GMC
model of \citet{goldreich74}, a simplified version of which is that
GMCs are spheres of gas of density $\rho$ that are in free-fall
collapse. The clouds reach a singularity in a time $t_{\rm
ff}(\rho)=[3\pi/(32 G \rho)]^{1/2}$, at which point their mass is
converted into stars. For a population of such clouds $\sfrff=1$,
which is inconsistent with the \citet{zuckerman74} result.

An important observational question, and a crucial test for theories
of how star formation is regulated, is to what densities and
length scales $\sfrff$ remains much smaller than unity. In other
words, is there a density at which something like the
\citet{goldreich74} free-fall collapse model becomes reasonable?
As an example, consider observing a star-forming region using
a molecular tracer sensitive to gas at densities of
$n_H\gtsim 10^{12}$ cm$^{-3}$, where $n_H$ is the number density of
hydrogen nuclei, to estimate the total mass of gas at such high
densities. This is larger than the mean density of ``cores'' seen both
observationally \citep[e.g.][]{barranco98} and in simulations of star
formation regulated by turbulence \citep[e.g.][]{jappsen05}, and is
roughly the density at which models of magnetically-regulated star
formation predict that gas will completely decouple from the magnetic
field and enter free-fall collapse \citep[e.g.][]{desch01}. At such
high densities protostellar outflows can probably stop at most half the
gas from reaching a star \citep{matzner00}, and the
thermal pressure in gas at $10^{12}$ cm$^{-3}$ is considerably
higher than the typical ram pressure of the turbulence in GMCs, so
gas at such high densities is largely impervious to external
perturbations. 

Thus, regardless of the model of star formation one adopts, one would
expect that almost all of the gas at densities $\gtsim 10^{12}$
cm$^{-3}$ is part of gravitationally-bound, collapsing objects that
have largely decoupled from the background turbulent flow. In the
absence of effective internal support or external disturbance, order
unity of the gas at such high densities is likely to be incorporated
into a star within one free-fall time. For this reason,
essentially all models of
star formation predict that the total mass of gas at densities
$\gtsim 10^{12}$ cm$^{-3}$, divided by the free-fall time of this gas,
should yield a value comparable to the total star formation rate in
the region over which the mass is measured. Instead of $\sfrff\sim
0.01$ as for GMCs, i.e. slow star formation, one would obtain
$\sfrff\sim 1$, i.e. rapid star formation.

However, different models make different predictions about the shape
of the curve of $\sfrff$
versus density in between $\sim 1\%$ at the characteristic density of
GMCs, $\sim 100$ cm$^{-3}$, and $\sim 1$ at a density $\gtsim 10^{12}$
cm$^{-3}$. These different predictions correspond to different models
of the physical scale at which gas both decouples from the background
flow and ceases to be supported by internal feedback mechanisms, and
thus transitions from slow to rapid star formation. At one extreme, magnetic
regulation models such as those of \citet{desch01}, neglecting for the
moment turbulent enhancement of the ambipolar diffusion rate
\citep[e.g.][]{heitsch04}, predict that star formation only becomes
rapid once gas decouples from the magnetic field, a process that does
not even begin until densities of $\sim 3\times 10^{10}$ cm$^{-3}$. At
the other extreme, \citet{bonnell03} argue that star clusters form from
gas clumps at densities of $\sim 5\times 10^{4}$ cm$^{-3}$ that undergo
a free-fall collapse in which at least $\sim 30\%$ of their mass is
converted into stars \citep{kroupa01b}, so star formation should be
rapid at this density or higher. In this model, the decoupling scale
corresponds to the transition from globally unbound structures (GMCs)
to globally bound structures (protoclusters). Thus, extending the
\citet{zuckerman74} calculation of star formation rate divided by
free-fall time to higher densities, in hopes of identifying a scale at
which there is a transition from slow to rapid star formation,
provides a means of distinguishing between models of how star
formation is regulated.

It is important at this point to differentiate the concepts of the
\textit{rate} and \textit{efficiency} of star formation, and the
\textit{lifetime} of star-forming clouds. Unfortunately these terms
are often confused in the literature, and there are no standard
definitions, so we describe here the definitions we use in this
paper. The star formation rate $\dot{M}_*$ is the easiest to
define, since it is simply the instantaneous conversion rate of gas
into stars within some volume. If we pick a density threshold $\rho$,
we can then define the dimensionless star formation rate per free-fall
time for the gas above that density threshold, $\sfrff =
\dot{M}_*/[M(>\rho)/t_{\rm ff}(\rho)]$, where $M(>\rho)$ is the mass
of material inside the volume of density $\rho$ or greater. In
contrast, the lifetime $t_{\rm cl}(\rho)$ of star-forming cloud is
somewhat more ambiguous. We take it to mean the total duration during
which a cloud is visible in a tracer sensitive to densities of
$\rho$ or more. Note that our definition neglects the complication
that something that starts as a single cloud may in the course
of its evolution break up
into multiple pieces, so the visible lifetime and the
dynamical lifetime may be different. Finally, by the efficiency
$\epsilon(\rho)$ we mean the fraction of gas mass $M(>\rho)$ that is
converted into stars by a cloud over its lifetime $t_{\rm
cl}(\rho)$. Again, we neglect the complication that clouds are not
closed boxes, so $M(>\rho)$ is likely to be time-dependent due to
continuing accretion or mass loss. Ideally $\epsilon(\rho)$
should be computed using a Lagrangian definition of the cloud mass,
i.e.\ all the mass that reached density $\rho$ or more at some
point. However, this is generally not a direct observable.

Roughly speaking, the rate,
efficiency, and lifetime are related by $\dot{M}_* \approx \epsilon(\rho)
M(>\rho)/t_{\rm cloud}(\rho)$. Consequently, one can define a
rough time-averaged value for $\sfrff$ in a cloud
$\left<\sfrff(\rho)\right> \approx
\epsilon(\rho)/[t_{\rm cl}(\rho)/t_{\rm ff}(\rho)]$, i.e.\ the mass
fraction converted into stars divided by the cloud lifetime in
free-fall times. One could also describe
$\left<\sfrff\right>$ as being the efficiency per free-fall time,
since it measures a fraction of mass converted into stars. However, we
refer to it as a rate because it is measured in amount per unit time,
whereas efficiency in the literature most commonly refers to the total
fraction of mass converted into stars, not the amount per unit
time. Moreover, since the time-averaged definition is ambiguous anyway
due to uncertainties in exactly what is meant by the efficiency and
the lifetime, we will generally use the instantaneous value of
$\sfrff$, which is well defined and, as we show, directly observable.

The reason for making all these definitions explicit it to point out
that observational constraints on $\sfrff$ by themselves do not
directly constraint $t_{\rm cl}$ or $\epsilon$, and vice
versa. Numerous authors have used various observational techniques
to estimate $t_{\rm cl}$ in clusters and GMCs
\citep[e.g.][]{elmegreen00, hartmann01, tassis04, mouschovias05,
tan06a, ballesterosparedes06, blitz06a}.  However, with the exception
of \S~\ref{clustertime}, we will not discuss the question of cloud
lifetime at all in this paper.
When we refer to rapid versus slow star formation, what we mean is
$\sfrff\sim 1$ or $\ll 1$, not is $t_{\rm cl}\sim t_{\rm ff}$ or $\gg
t_{\rm ff}$. These questions are conceptually distinct. As an example,
note that in the \citet{goldreich74} picture of GMCs as collapsing
spheres, the evolution time scale is always the cloud free-fall time,
$t_{\rm cl} \sim \tff$. However, the model still gives $\sfrff=1$, and
we would therefore describe it as a rapid star formation. In contrast,
\citet{ballesterosparedes06} argue that star formation lasts only one
crossing time of a molecular cloud, so again $t_{\rm cl}\sim \tff$,
but that during this time only $\sim 1\%$ of the mass turns into
stars. We would describe this as slow star formation, since
$\sfrff\sim 0.01$, even though the cloud evolution time is similar to
that in the \citeauthor{goldreich74} model. The
\citeauthor{zuckerman74} argument rules out the
\citeauthor{goldreich74} free-fall collapse model for GMCs, but is
consistent with the \citeauthor{ballesterosparedes06}
model. Thus, we emphasize that \textit{in this paper we remain
completely agnostic on the question of molecular cloud lifetime}.

In this paper we consider star formation in several classes of object.
Infrared dark clouds (IRDCs) are regions of high extinction seen in
absorption against the Galactic infrared background \citep{egan98,
carey00, simon06}. IRDCs are clearly associated with star formation,
and in at least some cases IRDCs have massive protostars
embedded within them \citep{rathborne05}. Several authors have
suggested \citep[e.g.][]{menten05,tan05b,rathborne06a}
that IRDCs are the progenitors of star clusters. Within IRDCs, at
still higher densities, are dense molecular clumps. These objects may
be observed in a variety of molecular transitions with high critical
densities, and we consider two here: HCN(1-0)
\citep{gao04b,gao04a,wu05} and CS(5-4)
\citep{plume97,shirley03}. Molecular clumps seen in these two
transitions are often associated with water masers and other signs
of massive, clustered star formation. Our goal is to determine
$\sfrff$ for each of these increasingly dense gas tracers. We also
determine this quantity for the Orion Nebula Cluster using a
completely different method, which provides an independent check on
our estimates.

The remainder of this paper proceeds as follows: in \S~\ref{sfrest} we
use a variety of observations to derive $\sfrff$ for our objects
and construct a plot of $\sfrff$ versus characteristic
density to search for signs of a transition from slow to fast star
formation. In \S~\ref{implications}, we compare our results to
theoretical models for the regulation of star formation, and also
point out some implications for the process of star cluster
formation. Finally, in \S~\ref{conclusions} we summarize our
conclusions, and suggest directions for future work.

\section{Estimates of $\sfrff$}
\label{sfrest}

Let $\sfr$ be the total star formation rate in a galaxy, and
consider star formation occuring in objects of class X. Let $\fx$ be
the fraction of galactic star formation that occurs in these objects,
$\mx$ be their total mass in the galaxy, and $\tffx$ be their
typical free-fall time. Then
\begin{equation}
\label{sfrffeqn}
\sfrffx = \frac{\fx \sfr \tffx}{\mx}.
\end{equation}
In this section we use observations to estimate all the factors on the
right hand side, and thus determine the dimensionless star formation
rate $\sfrffx$ in several classes of object. We emphasize that, as we
vary the typical density of the objects we consider, both $\tffx$
and $\mx$ will vary -- as we go to denser and denser clouds, the
free-fall time will decline, pushing $\sfrffx$ downward, but the
amount of mass in objects of that density will also decrease, pushing
$\sfrff$ back up. Ultimately, we wish to determine which of these two
effects dominate, or whether they roughly balance.

In addition to determining a best value for $\sfrffx$, we also attempt
to make a rough estimate our uncertainties by estimating the uncertainty on
each parameter in equation (\ref{sfrffeqn}). We do this from data where
possible and from more general considerations where not, and we add
the uncertainty factors in quadrature. This method of combining errors
is formally appropriate only if the errors are independent and
Gaussian-distributed, neither of which are likely to be completely
true, but the procedure does give us a sense of by how much our
estimates could vary.

\subsection{Star Formation Rates}

The unknown on the right hand side of (\ref{sfrffeqn}) that has been
studied most heavily is the star formation rate. Numerous authors have
discussed methods of inferring the star formation rate from various
observables \citep[e.g.][and references
therein]{kennicutt98b,iglesiasparamo06}, so here we only summarize
aspects of this discussion that we will apply directly in what
follows.

\subsubsection{The Milky Way}

In the Milky Way, \citet{mckee97} find $\sfr\approx 3$ $\msun$
yr$^{-1}$ based on catalogs of Galactic HII regions. Star formation is
distributed in an exponential disk with scale radius $H_R=3.3$ kpc and
sharp cutoffs at 3 kpc and 11 kpc in Galactocentric radius. For this
distribution, approximately $2/3$ of Galactic star formation occurs
within the solar circle. The dominant uncertainty in this estimate,
roughly 0.3 dex \citep{kennicutt98b}, is the shape of the stellar
initial mass function (IMF), since
HII regions only trace the massive stellar population and one must
extrapolate to estimate the total mass. Estimates of the star
formation rate based on chemical modelling \citep{prantzos95} also
give star formation rates consistent with 3 $\msun$ yr$^{-1}$ to
within a factor of 2.

\subsubsection{Extragalactic Far-Infrared Observations}
\label{irsfsection}

For external galaxies, a commonly-used tracer of star formation is
far-infrared (FIR) light. While FIR is usually not the preferred tracer of
star formation in normal spiral galaxies like the Milky Way, it is
detectable over a very wide range of sources, from ultraluminous
infrared galaxies (ULIRGs) to individual cluster-forming clumps in the
Milky Way, making it possible to look for correlations over a wide range of
luminosities \citep[e.g.][]{gao04a,wu05}. \citet{kennicutt98b} finds
that in optically-thick starbursts the IR luminosity and
star-formation rate are related by
\begin{equation}
\label{irsfrconversion}
\sfr/\left(\msun\mbox{ yr}^{-1}\right) = 2\times 10^{-10} (L_{\rm IR}/\lsun).
\end{equation}
The dominant uncertainty is the age of the stellar population, but
as we discuss in more detail below, for extragalactic sources this is
only $\sim 30\%$. The uncertainty is considerably larger in normal spiral
galaxies, where old stellar populations contribute a significant
luminosity, and where only a fraction of the light is reprocessed into
infrared. Based on comparisons of multiple tracers of the star
formation rate in a range of galaxy types, \citet{iglesiasparamo06}
find that (\ref{irsfrconversion}) probably underestimates the star
formation rate in normal spirals. We instead adopt
\begin{equation}
\label{irsfrconversion1}
\sfr/\left(\msun\mbox{ yr}^{-1}\right) = 4\times 10^{-10} (L_{\rm IR}/\lsun)
\end{equation}
in normal spirals, which is approximately consistent with the mean in the
sample of \citet{iglesiasparamo06} for low star-formation rate ($\sfr
\ltsim 10$ $\msun$ yr$^{-1}$) systems. There is, however, a much larger
scatter in this relation than in the corresponding relation for
starburst systems.

\subsubsection{Galactic Far-Infrared Observations}
\label{galfir}

Several authors \citep[e.g.][]{plume97,mueller02,shirley03,wu05}
use infrared luminosities to estimate the star formation rates within
individual cluster-forming gas clumps in the Milky Way. These objects have very high
surface densities, so essentially all the light is
reprocessed into IR and the uncertainty for spiral galaxies does not
apply. However, we will not directly compare to these data because
there is a much larger uncertainty arising from the
age of the stellar population. In systems $\ltsim 3$ Myr old no stars
have disappeared through supernovae, so the massive stellar population
cannot yet have reached equilibrium between formation and destruction.

\begin{figure}
\plotone{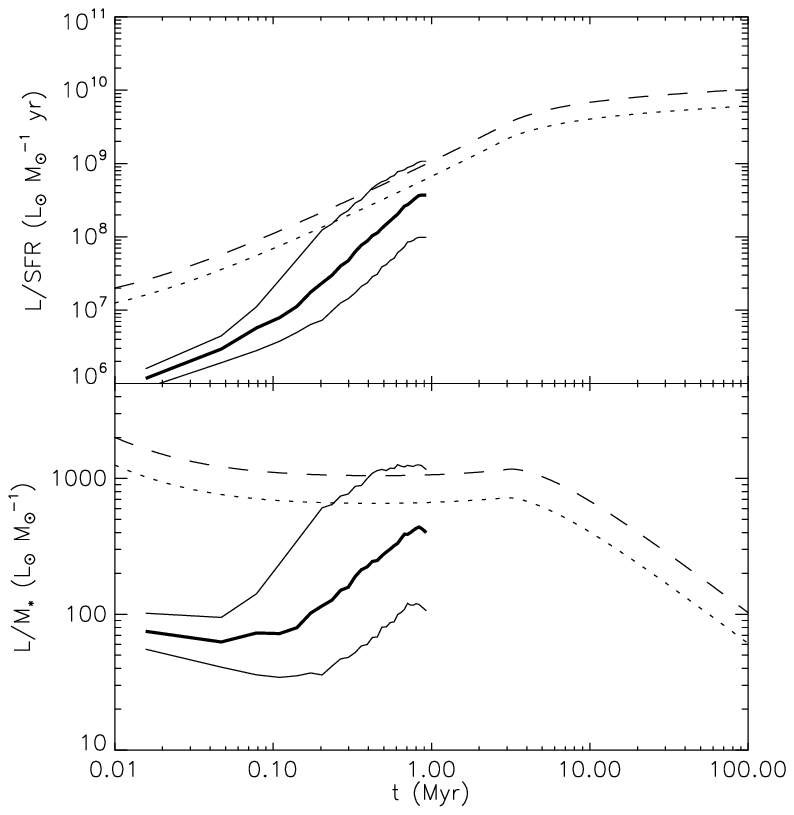}
%\plotone{f1.pdf}
\caption{
\label{lt}
Luminosity to star formation rate ratio (\textit{upper panel}) and
luminosity to stellar mass ratio (\textit{lower panel}) versus time
for a stellar population forming at a constant rate. We show the
following computations: using starburst99 version 5.0
\citep{leitherer99,vazquez05} with the default parameters
(\textit{dashed lines}); using starburst99 with an IMF that has a
slope of $-2.35$ from 0.1 $\msun$ to 120 $\msun$ (\textit{dotted
lines}); and the median (\textit{thick solid lines}) and $1\sigma$
upper and lower limits (\textit{thin
solid lines}) for 400 clusters simulations computed with the model of
\citet{tan02}. Further details are given in \S~\ref{galfir}.
}
\end{figure}

To study the magnitude of the uncertainty that this effect induces, we
use starburst99 version 5.0 \citep{leitherer99,vazquez05} to compute
the evolution of light to star formation rate and light to mass ratios
for systems with constant star formation rates. Figure \ref{lt}
shows the results. The dashed lines show models using the
\citet{kroupa02} IMF, while the dotted lines use an IMF with the
\citet{salpeter55} slope of $\alpha=-2.35$ from 0.1 $\msun$ to 120
$\msun$; for all other starburst99 parameters, we use the defaults.
As the plot shows, at ages of a few Myr or more, the luminosity per
unit star formation rate is quite insensitive to both age and IMF, so
luminosity is a good indicator of star formation rate. However, at
younger ages the luminosity in the starburst99 calculation traces
total stellar mass more closely than star formation rate, so the light
is not a good indicator of star formation rate.

One might think it possible to break this degeneracy by independently
determining the age of the stellar population, and then using the
luminosity to infer the stellar mass and thus the star formation
rate. However, a more detailed treatment of very young systems than
starburst99 provides shows that infrared observations alone are not
sufficient to constrain the stellar mass in systems $\ltsim 1$ Myr
old. In such young systems, a significant fraction of the stars may
not yet have contracted to the main sequence, in which case they will
be more luminous than equal mass stars on the main sequence would
be. Accretion luminosity may further enhance the radiative output above
what would be found for a non-accreting population of the same
mass. There are also factors that reduce the luminous output compared
to an older population of the same mass. Massive stars require $\sim
0.1$ Myr to assemble \citep{mckee02, mckee03}, so systems younger than
this will be missing their contribution to the light. In systems of
$\sim 1000$ $\msun$ or less, poor sampling of the massive tail of the
stellar IMF produces a large scatter, and causes the median system
to be less luminous per unit mass or per unit star formation rate than
a larger population would be.

We explore how these effects change the light output of a young cluster
using the model of \citet{tan02}. In Figure \ref{lt}, the solid lines
show the light output from a simulated cluster with a final stellar mass of
$1000$ $\msun$ forming at a constant star formation rate of $1.08\times
10^{-3}$ $\msun$ yr$^{-1}$. This model uses the same IMF we use for the
Salpeter-slope starburst99 calculation (the dotted line in the
figure), but it includes accretion 
luminosity, pre-main sequence evolution, finite star formation times,
and discrete sampling of the IMF. It does not include any post-main
sequence evolution. The thick central line shows the median of 400
runs with different samplings of the IMF, and the thin lines above and
below it show the tracks that bound $68\%$ of the runs. As
the plots show, the median cluster $\ltsim 1$ Myr in age will be much less
luminous per unit star formation rate than a galactic stellar population
that is $\gtsim 10$ Myr old and contains enough stars to fully
sample the IMF, but the spread can be an order of magnitude for a 1000
$\msun$ cluster. As the cluster mass increases the effects of discrete
sampling decrease, causing the luminosity spread to decrease and the
median to rise. Even for
clusters massive enough to sample the full IMF, though, neither the
light to mass ratio nor the light to star formation rate ratio stay
constant at ages $\ltsim 1$ Myr, so light is a poor indicator of
either stellar mass or star formation rate.

While we cannot use infrared luminosity to study the star formation
rate in Galactic star-forming gas clumps directly, there is an
observational correlation between infrared and molecular luminosity
for such objects \citep{wu05} from which we can learn a great deal. 
We discuss how to interpret this correlation in light of our results
in \S~\ref{clustertime}.

Finally, note that there are few star clusters where one can estimate
the star formation rate directly by placing a large number of stars on
the HR diagram and using pre-main sequence tracks to estimate the
cluster mass and age spread
\citep[e.g.][]{palla99,palla00,huff06}. Since this requires luminosity
and temperature determinations for many stars, it is posssible only in
systems without too much extinction, which limits this technique to
low density regions (e.g. Taurus) or somewhat older regions where most
of the initial gas is gone (e.g. the Orion Nebula Cluster, ONC). We
discuss the ONC in more detail in \S~\ref{ONCsec}, and also refer
readers to \citet{tan06a} for a detailed discussion of other
techniques by which one may estimate the formation times of star
clusters.

\subsection{Star Formation in Infrared Dark Clouds}
\label{IRDCsfr}

As discussed above, IRDCs are likely to be the progenitors of star
clusters. \citet{simon06} catalog the infrared dark clouds in the
inner Galaxy. This enables us to compute the star formation rate per free
fall time in IRDCs by comparing to the total star formation rate in
the inner Galaxy, roughly $2$ $\msun$ yr$^{-1}$ with a factor of 2
uncertainty. We should note that our calculation of the star formation
rate in IRDCs is entirely different from the one given
\citet{rathborne06a}. Their estimate assumes an unknown time scale
and efficiency of star formation. The calculation we give here does
not rely on any such assumptions.

First we must estimate the fraction of star formation that occurs in
IRDCs. \citet{lada03} find that $70-90\%$ of Galactic
star formation occurs in clusters. If a cloud were visible during the
entire star formation process as an IRDC, this would imply $\firdc
\approx 0.8$. However, clouds that have too many embedded protostars
will not be infrared dark, and will therefore cease to be visible as
IRDCs. This probably does not occur until most of the stars have
formed, though, so we adopt $\firdc = 0.8$ as a reasonable guess,
while acknowledging that it could be a bit smaller. Being slighly more
conservative about our uncertainty than \citeauthor{lada03}, we adopt 
$\firdc = 0.6-1$ as the plausible range of possibilities,
i.e.\ anywhere from 60\% to 100\% of star formation inside the solar
circle takes place in IRDCs, corresponding to a factor of 1.25
uncertainty on our estimate of $\firdc=0.8$. Note that $\firdc=1$
gives the largest possible value of $\sfrff$ for IRDCs, so any
uncertainties that move $\firdc$ out of our plausible range can only
make $\sfrff$ smaller, not larger.

\citet{rathborne06a} find that the total mass of IRDCs in
the inner Galaxy is $\mirdc \approx 10^8$ $\msun$ based on the measured 
properties of a sub-sample of IRDCs observed in 1.2 mm continuum
emission, and an
estimate of the detection efficiency for the MSX IRDC survey
\citep{simon06}. This mass estimate is probably uncertain by a factor
of several, because it is unclear how representative the clouds
surveyed by \citet{rathborne06a} are of the entire IRDC
population. The sample consists of the darkest clouds (darkness
measures a combination of column density and degree of background
illumination) from a sample with known kinematic distances. This
selection introduces an unknown bias in the mass estimate, so we
consider the IRDC mass estimate to be uncertain by factors of a few.

\begin{figure}
\plotone{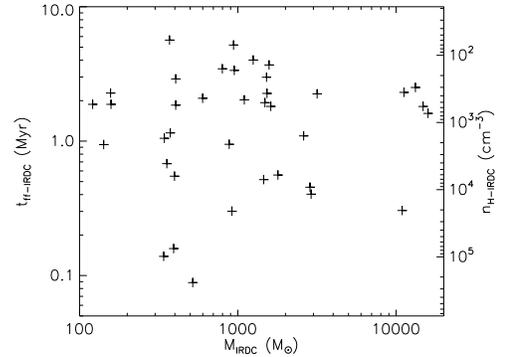}
%\plotone{f2.pdf}
\caption{
\label{tffirdc}
Free-fall time and density versus mass for the 38 IRDCs in the sample
of \citet{rathborne06a}.
}
\end{figure}

The free-fall time in IRDCs is $\tff=[3\pi/(32 G
\rho)]^{1/2}$, where $\rho$ is the mean density. For the 38 IRDCs
in the sample of \citet{rathborne06a}, we define an effective radius
$r=(A/\pi)^{1/2} D$, where $A$ is the angular area within the
$2\sigma$ detection threshold of the cloud taken from the catalog of
\citet{simon06}, and $D$ is the distance estimate taken from Rathborne
et al. We take the mean density to be $\rho=3M/(4\pi r^3)$. This procedure
is fairly uncertain, since the location of the $2\sigma$ contour
depends on the background emission, and the morphology is filamentary
rather than round for a significant minority of clouds. Nonetheless,
we can make a rough estimate for $\rho$ and $\tff$, and check for
any systematic variations with IRDC size. We plot the derived
free-fall times and densities in Figure \ref{tffirdc}. As the plot
shows, there is a spread of a factor of $\sim 3$, but no clear
systematic trend. We adopt as our characteristic free-fall time the
mass-weighted harmonic mean 
\begin{equation}
\label{harmonicmean}
\tffirdc\equiv \frac{\left\langle M\right\rangle}{\left\langle M/\tff
\right\rangle}
=0.9\mbox{ Myr},
\end{equation}
corresponding to a number density of hydrogen nuclei $n_H=2\times
10^3$ cm$^{-3}$. The range of free-fall times covering the central
67\% of the IRDCs is $0.45-3.0$ Myr, so the uncertainty is roughly a
factor of 2.6. Similarly, the central 67\% of IRDC densities covers a
range of $210-9200$ cm$^{-3}$, an uncertainty of a factor of 6.6.

We now have all the necessary figures to plug into equation
(\ref{sfrffeqn}): the fraction of star formation inside the solar
circle taking place in IRDCs is $\firdc=0.8$, with a $25\%$
uncertainty; the total star formation rate inside the solar circle is
$2$ $\msun$ yr$^{-1}$, with a factor of 2 uncertainty;
the free-fall time in IRDCs is $0.9$ Myr, with a factor of 4
uncertainty; the total mass of IRDCs in the inner Galaxy is $10^8$
$\msun$, with a hard-to-quantify uncertainty, which we take to be a
factor of 3. Combining these estimates gives a fiducial value $\sfrffirdc
\approx 0.014$. Combining the errors as described in \S~\ref{sfrest}
gives an uncertainty of a factor of 4.6.

\subsection{Star Formation in Gas Traced by HCN}

The molecular transition HCN(1-0) has a critical density of
$n_H=6\times 10^{4}/\tau$ cm$^{-3}$ \citep[c.f.][who give the critical
density in terms of number density of hydrogen molecules rather than
hydrogen nuclei]{gao04b}, where $\tau$ is the line-center optical
depth in the escape probability approximation, and is therefore an excellent
tracer of the dense molecular regions associated with star formation.
\citet{gao04b,gao04a} observe a large sample of normal spiral
galaxies and luminous and ultraluminous infrared galaxies in
HCN(1-0) and show that the HCN luminosity correlates well with
the IR luminosity, following
\begin{equation}
\frac{L_{\rm IR}}{\lsun} = 911 \pm 227
\frac{L_{\rm HCN}}{\mbox{K km s}^{-1}\mbox{ pc}^{-2}},
\end{equation}
where the figure quoted is the mean plus or minus the standard
deviation. Galactic cores with luminosities above $10^{4.5}$ $\lsun$
observed in HCN by \citet{wu05} show a similar correlation, with
almost exactly the same mean, although the median is lower than the
median of the extragalactic sample by a factor of 2.5. Since infrared
luminosity is a tracer of star formation rate, and HCN(1-0) is a
tracer of molecular gas, this relation is a direct measure of the star
formation rate per unit mass in gas of densities traced by HCN, and
thus of $\sfrff$ in that gas. We use the difference in medians between
the Galactic and extragalactic samples as an estimate of the
uncertainty factor in the IR-HCN correlation, although we note that
this is almost certainly an overestimate for the extragalactic sample
because that data set is much more tightly correlated than the
Galactic sample.

To derive the total gas mass that corresponds to a given luminosity
$L_{\rm HCN}$ in the HCN(1-0) line,  \citet{wu05} observe in HCN(1-0)
a sample of star-forming clumps with known virial masses determined
from optically-thin C$^{34}$S emission. (Other methods of estimating
the mass of these objects that do not assume virial balance, and make
other assumptions about the emission, e.g. \citealt{plume97},
\citealt{mueller02}, \citealt{shirley03}, give masses that are
comparable to or better than the uncertainty in the mass estimate we
adopt below.) They find a median ratio
$\mhcn/\msun = 6 \,L_{\rm HCN}/\left(\mbox{K km s}^{-1}\mbox{
pc}^2\right)$, and a mean of 11. \citet{gao04b} perform radiative
transfer calculations under a variety of assumptions about the
temperature and distribution of the HCN-emitting gas, and arrive at
conversion factors of $10-25$. For normal spiral galaxies, we adopt
a conversion of $\mhcn/\msun = 11 \,L_{\rm HCN}/\left(\mbox{K km
s}^{-1}\mbox{ pc}^2\right)$, the mean value measured by \citet{wu05}.
To be conservative we estimate that the conversion is uncertain by a
factor of 3, corresponding to a range that is larger than the range
of $6-25$ spanned by the various estimates.

Recent HCO$^+$ observations suggest that chemical changes triggered by
X-rays may enhance the HCN abundance in galaxies with an AGN
\citep{graciacarpio06}. This reduces the mass estimate for a given HCN
luminosity by a factor of $\sim 2$ in starbursts, most of which have at
least a small AGN. For starbursts we therefore modify our light to
mass conversion to $\mhcn/\msun = 5.5 \,L_{\rm HCN}/\left(\mbox{K km
s}^{-1}\mbox{ pc}^2\right)$, again with roughly a factor of 3
uncertainty. 

Combining this with the conversion from IR luminosity to star
formation rate discussed in \S~\ref{irsfsection} for both starburst
galaxies and normal spirals gives $\sfr/\mhcn \approx (30\mbox{
Myr})^{-1}$. Adding the uncertainties of a factor of 2.5 on the IR-HCN
correlation, 3 on the HCN to mass conversion, and 1.3 on the IR to star
formation rate conversion gives a combined uncertainty of a factor of
4.1.

The free-fall time depends on the mean density, for which a rough
guess based on the optical thickness of the HCN emission is
$n_H\sim 6\times 10^4$ cm$^{-3}$. The true density of the HCN-emitting
gas could be lower or higher depending on optical depth and
beam-filling effects, as appears to be the case with CS(5-4) emission
(see the discussion below in \S~\ref{cssection}). However, this is
unlikely because the mass to luminosity ratio for HCN(1-0) inferred
from radiative transfer calculations is very close to that inferred
from correlation of HCN luminosity with virial masses for objects seen
in the Galaxy. This suggests that the HCN emission for Galactic HCN
clumps is close to beam-filling and is not extremely self-absorbed,
so the density of the objects being observed is unlikely to be very
different than $n_H\sim 6\times 10^4$ cm$^{-3}$. If anything, the
observations of \citet{wu05} suggest that the mean density may be
slightly larger than $n_H\approx 6\times 10^4$ cm$^{-3}$ (also J. Wu,
2006, private communication). We adopt a mean density of $n_H\sim
6\times 10^4$ cm$^{-3}$ as a fiducial value, giving
$\tffhcn\approx 0.18$ Myr, and estimate that the uncertainty is no
more than a factor of 2, corresponding to a factor of 4 change in the
density.

Finally, since the extra-Galactic
observations include all HCN(1-0) emission from the target galaxy, and
all star-formation occurs at densities high enough to be traced by HCN
emission, we set $\fhcn=1$. 

Using our fiducial estimates for $\sfr/\mhcn$, $\fhcn$, and $\tffhcn$
in (\ref{sfrffeqn}) gives $\sfrffhcn = 0.0058$. The factor of 4.1
uncertainty in our value of $\sfr/\mhcn$, combined with the factor of
2 uncertainty in our free-fall time, gives a complete uncertainty of a
factor of 4.6.

\subsection{Star Formation in Gas Traced by CS}
\label{cssection}

The CS(5-4) line has a critical density of $n_H \approx 1.5\times 10^6/\tau$
cm$^{-3}$, and thus traces gas at even higher densities than
HCN(1-0). \citet{plume97} and \citet{shirley03} survey in CS emission
lines a sample of regions selected from sources reported in the
Arcetri H$_2$O maser catalog \citep{valdettaro01} that are thought
based on IRAS colors to be associated with star formation. From the
mean CS(5-4) luminosity of their targets and the sky coverage fraction
of the maser catalog, they estimate that the total CS(5-4) luminosity
of the Galaxy is at least $L_{\rm CS} \approx 20$ $\lsun$. This is a
lower limit only, because the maser catalog may not
be complete over the region of sky it covers, there is probably at
least some Galactic CS(5-4) emission that is not correlated with water
masers, and \citet{shirley03} only detect 75\% of their
targets in CS(5-4), while they detect 90\% in all the transitions for
which they look. Nonetheless, this lower limit on luminosity gives a
lower limit on the amount of gas in the Galaxy that is sufficiently
dense to produce CS(5-4) emission, which we can in turn use in
equation (\ref{sfrffeqn}) to obtain an upper limit on $\sfrff$ at CS
densities.

\begin{figure}
\plotone{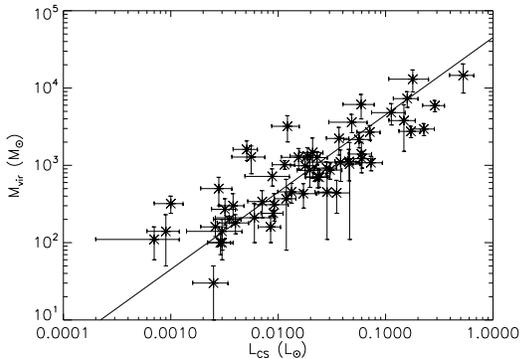}
%\plotone{f3.pdf}
\caption{
\label{mvirlcs}
Virial mass $M_{\rm vir}$ versus luminosity in the CS(5-4) line
$L_{\rm CS}$ (\textit{asterisks with error bars}) for objects in the
\citet{shirley03} sample. We also show the best-fit linear correlation
(\textit{line}).
}
\end{figure}

First we must estimate the CS(5-4) ``X'' factor to convert the
luminosity to a mass. Like HCN(1-0), star-forming clumps are usually
optically thick in CS(5-4), so it is reasonable to expect that such an
X factor might exist. Figure \ref{mvirlcs} shows the virial mass $M_{\rm
vir}$ (determined from optically-thin C$^{34}$S emission) versus
$L_{\rm CS}$ for the 57 objects in the \citet{shirley03} sample. The
correlation is well fit by the line $\mcs/\msun = 4.5\times 10^{4}
\, L_{\rm CS}/\lsun$ over more than two orders of magnitude in
mass. This implies a total Galactic mass in CS clumps of $\mcs \geq
9\times 10^5$ $\msun$. The uncertainty of the fit, done in the
logarithm of the data, is a factor of 3.3, and we adopt this as the
uncertainty in our luminosity to mass correlation. However, as we
discuss more below, the virial mass gives the lowest mass of the three
estimators tested by \citet{plume97}, by roughly a factor of 3, so the
true mass is much more likely to be larger than our value rather than
smaller.

Note that this mass estimate strongly suggests that the true CS
mass in the Milky Way is substantially larger. CS and HCN emission
often come from overlapping regions in the Milky Way \citep{wu05}, and
in such overlapping regions the area emitting CS is not vastly
smaller than the area emitting HCN. However, the total mass of CS
gas that our estimate suggests is smaller than the mass of
HCN-emitting gas that \citet{gao04b,gao04a} see in galaxies like the
Milky Way by $1-2$ orders of magnitude. The most likely
explanation is incompleteness of the CS survey.

\begin{figure}
\plotone{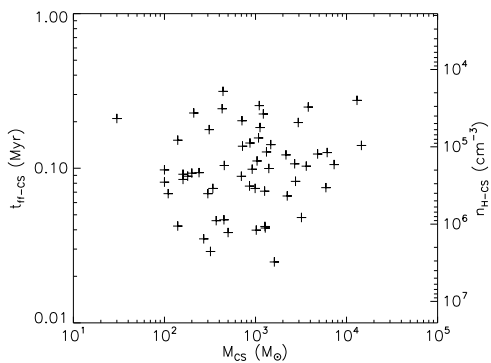}
%\plotone{f4.pdf}
\caption{
\label{tffcs}
Free-fall time and density versus mass for the 57 CS(5-4) clumps in
the sample of \citet{shirley03}.
}
\end{figure}

To estimate the free-fall time, we must know the gas
density. \citet{plume97}, based on radiative transfer modeling, find
a mean density of $n_H=1.6\times 10^6$ cm$^{-3}$, roughly the critical
density. However, this average is intensity-weighted, so it may
overestimate the true mean. \citet{plume97} and \citet{shirley03} find
that the virial mass for objects in their survey is systematically
smaller than the mass estimated by assuming that all the gas is at the
density inferred from the radiative transfer calculations. Based on
the difference in mass estimates, they conclude that the filling
factor of gas at densities of $n_H \sim 2 \times 10^6$ cm$^{-3}$ or
higher is typically $\sim 0.3-0.5$. Thus, the density of
$n_H=1.6\times 10^6$ cm$^{-3}$ probably represents an upper limit on
the true density. We therefore compute the mean density for the 57
clumps in the Shirley et al. catalog from the virial mass and
deconvolved half-peak radius (Shirley et al.'s $R_{\rm CS}$) to
compare to the result of the radiative transfer calculations. We show
our derived densities and the corresponding free-fall times in Figure
\ref{tffcs}. The mass-weighted harmonic mean free-fall time
for the CS clumps (computed from equation \ref{harmonicmean}) is
$\tffcs=0.10$ Myr, and we adopt this as our fiducial value. The range
$0.05-0.2$ Myr covers the central 67\% of the free-fall times, so we
take the uncertainty to be a factor of 2. The density corresponding to
our adopted mean free-fall time is $n_H=1.8\times 10^5$
cm$^{-3}$, and the range in free-fall times corresponds to a density
range of a factor of 1.4.

Finally, the estimated CS luminosity comes from sources associated
with water masers, which arise only in regions of massive,
clustered star formation. This implies a maximum of
$\fcs=0.8$. Alternately, we would obtain the same result by assuming
that all star-forming regions produce CS(5-4) emission and raising our
estimate of $\mcs$ by a factor of $1/0.8$ to account for the fraction
of CS(5-4) emission associated with water masers. In either case, our
value for either $\fcs$ or $\mcs$ is an upper limit because we do not
know what fraction of clustered star formation is associated with
water maser emission. Combining $\fcs$, $\mcs$, and $\tffcs$ in
equation (\ref{sfrffeqn}) gives $\sfrffcs \leq 0.27$, with an
uncertainty of a factor of 3.6. Again, it seems likely that the true
value is significantly lower than this upper limit.

\subsection{Star Formation in the Orion Nebula Cluster}
\label{ONCsec}

We can add one more point for a specific object. The Orion Nebula
Cluster (ONC) is the only star cluster
that has been studied well enough so that it is possible to make
reasonable estimates of the stellar mass, the formation time, and the
properties of the progenitor gas system, allowing direct determination
of $\sfrff$. This is quite useful because it provides an estimate of
$\sfrff$ that does not depend on conversions from luminosities to
masses or star formation rates, and thus is subject to completely
different systematic errors than the methods we have used thus far.

The total stellar mass of the ONC is 4600 $\msun$
\citep{hillenbrand98}. \citet{tan06a} analyze several lines of
evidence to conclude that formation of the ONC took place over at
least $3$ Myr, and possibly longer -- see \citet{huff06}. The age
estimate comes from a combination of pre-main sequence fitting, which
is fairly reliable for stars $\gtsim 2$ Myr in age, and from the age
of a dynamical ejection event. We refer readers to \citet{tan06a} for
the detailed evidence. The free-fall time in the current stellar
system is $\tff=0.5$ Myr, but it must have
been smaller in the progenitor gas system, since some of the mass has
been expelled by the Trapezium stars, and the cluster may be
expanding. Uncertainties in these processes lead to a range of mass
estimates for the progenitor, ranging from $\sim 15000$ $\msun$
if the ONC today is marginally unbound and expanding \citep{kroupa01b}
to $6700$ $\msun$ if the ONC is currently bound and non-expanding
\citep{huff06}, although the latter would imply an
extraordinarily high star formation efficiency. If we neglect the
possibility that the cluster has undergone significant expansion since
expelling its gas, this implies that the free-fall time in the
progenitor gas system was $\tff=0.3-0.4$ Myr, corresponding to a
density $n_H=1-2\times 10^4$ cm$^{-3}$. It seems unlikely that star
formation began when most of the gas in the ONC was spread over a much
larger region at lower density, both because there is no correlation
between the age and spatial distribution for stars in the ONC
\citep{huff06} and because there must have been a system sufficiently
dense to lead to dynamical interactions and ejection of massive stars
2.5 Myr ago \citep{hoogerwerf01, tan06a}.

 Combining the initial
gas mass, final stellar mass, and initial free-fall time implies that
in Orion $\sfrff = 0.03-0.09$. The true value is likely to be towards
the low end of this range, since even if the ONC is bound it has
almost certainly undergone some expansion from its original gaseous
state, meaning that the free-fall time we have adopted is probably too
large.

\subsection{Summary of Observations}
\label{obssummary}

\begin{figure*}
\plotone{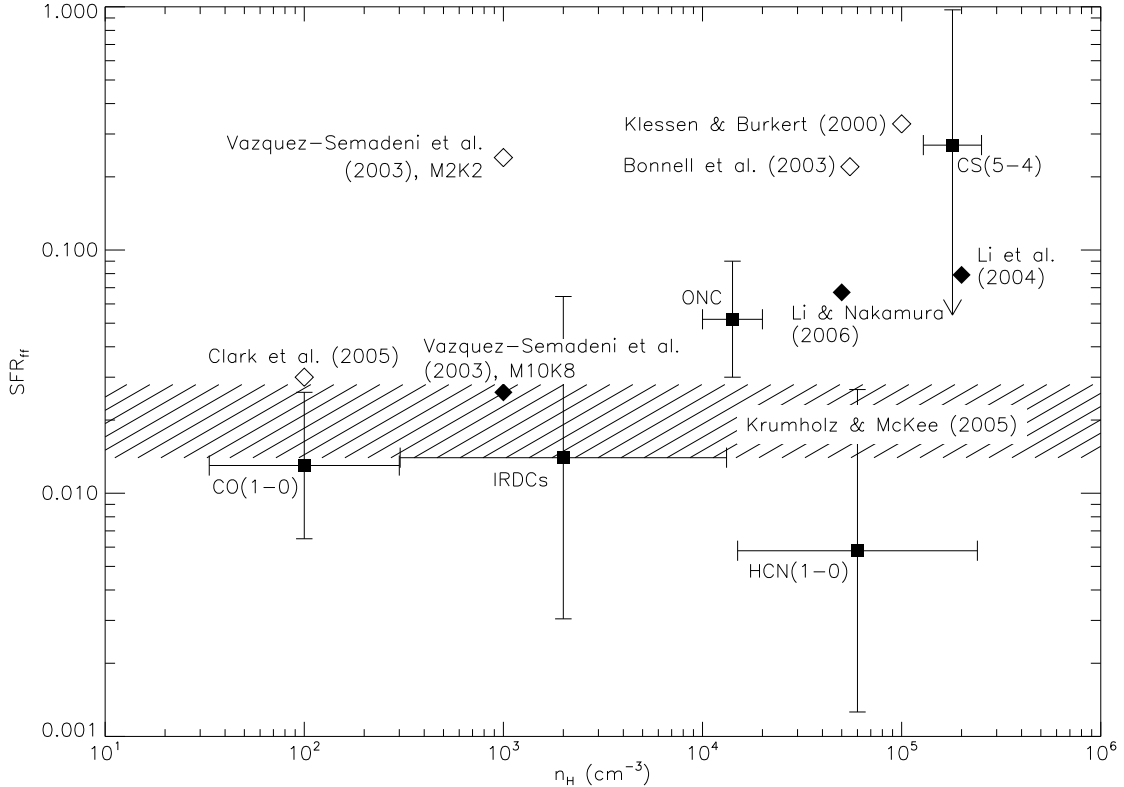}
%\plotone{f5.pdf}
\caption{
\label{sfrffn}
Estimates of the dimensionless star formation rate $\sfrff$ in objects
of characteristic density $n_H$. We show observational estimates
(\textit{filled squares with error bars}), an observational upper
limit (\textit{filled square with arrow}), simulations with weak or 
decaying turbulence (\textit{open diamonds}), simulations with strong,
driven turbulence (\textit{filled diamonds}), and the prediction of an
analytic model (\textit{hatched region}). For the observational
points, we label each point by the tracer used to estimate that
point. We label the simulation points and the analytic model region by
the reference for the simulation or model. The simulations of
\citet{klessen00b}, \citet{vazquezsemadeni03}, and \citet{li04} are
dimensionless, so the values of $n_H$ for them are arbitrary. We have
adopted the authors' suggested scale value of $n_H$ where one is
given, or an arbitrary value otherwise. For the simulations of
\citeauthor{vazquezsemadeni03}, we show a weakly-driven model (M2K2,
meaning Mach number 2, driving wavelength of $1/2$ the box size), and
a strongly-driven one (M10K8), and we derive values of $\sfrff$ for these
models via the method described in \citet{krumholz05c}.
The region indicated for the \citeauthor{krumholz05c} model
corresponds to the parameter range $\avir=1-2$, $\calm=20-40$.
}
\end{figure*}

We summarize our results by plotting the dimensionless star
formation rate $\sfrff$ versus characteristic density in Figure
\ref{sfrffn}. In addition to our points for CS clumps, HCN clumps,
IRDCs, and the ONC, we can add a point for GMCs as a whole, in which
the typical density and free-fall time are $n=100$ cm$^{-3}$ and
$\tff=4.4$ Myr \citep{mckee99a}. The total mass of GMCs in the Galaxy
is roughly $10^9$ $\msun$ \citep{bronfman00}, so using the same
argument as in \S~\ref{IRDCsfr} gives $\sfrff \approx
0.013$. \citet{krumholz05c} give a much more detailed calculation of
this value, but for simplicity we adopt rough numbers here. Since the
GMC population in the Milky Way is reasonably well characterized, the
uncertainty in $\sfrff$ is dominated by uncertainty in the star
formation rate, so it is about a factor of 2. The range
of densities covered by star-forming GMCs is roughly a factor of 3, so
we take  this range and the range in $\sfrff$ to roughly define our
uncertainty for GMCs.

\section{Implications and Comparison to Theory}
\label{implications}

The star formation rate per free-fall time is a key prediction of any
model for a physical mechanism that regulates the star formation
rate. The most important result summarized by Figure \ref{sfrffn},
which any successful theory must be able to explain, is that there is
no evidence for a transition to rapid star formation out to densities
of $n\sim 10^5$ cm$^{-3}$. Even given the large uncertainties in the
data, this conclusion is quite firm. If any of the objects we have
analyzed were bound, detatched from the large-scale turbulent flow,
and collapsing without significant impedance from feedback, we would
expect $\sfrff \sim 1$, while even the top of the uncertainty range
for the data point with the highest value of $\sfrff$ is below
0.1. Only the CS data is consistent with a transition for rapid star
formation, and this is an upper limit for which there is strong
evidence to suggest that the true value is significantly lower.

A second conclusion, which is much more tentative, is that we do not
see any significant change in $\sfrff$ with density at all. Given the
uncertainties, obviously we are insensitive to changes in $\sfrff$
smaller than roughly an order of magnitude, and the somewhat higher
point for the ONC suggests that it is possible there is some modest
rise in $\sfrff$ in star clusters. Any such rise with density, though,
must be fairly small over the range of density we have explored. In
summary, the
data are \textit{consistent} with $\sfrff$ of a few percent regardless
of density, and \textit{require} $\sfrff\ltsim 0.1$ to densities of at
least $6\times 10^4$ cm$^{-3}$. The characteristic scale at which one
transitions from slow to fast star formation must lie at even higher
densities. Here we investigate how well various models of star
formation explain this result, and discuss a few of its implications.

\subsection{Unbound GMC / Collapsing Cluster Models}

One proposed explanation for the low star formation rate in molecular
clouds is that GMCs are not bound by gravity
\citep{maclow04,clark04,clark05,dobbs06a}. In this model,
GMCs are transient over-densities created by turbulence or spiral
shocks in the atomic ISM. Since they are confined by ram pressure,
they re-expand and disperse in roughly one dynamical time. The star
formation rate appears low because only a small fraction of the mass is
gravitationally bound and can collapse, but that mass which can
collapse will form stars rapidly. Feedback is not effective at slowing
collapse or establishing an equilibrium, and only stops the star
formation process at its end by rapidly unbinding
the cluster \citep{bonnell06c}. Star formation in protoclusters
is rapid but is halted after a short period by feedback. Thus, the
star formation rate is high, even if the duration is short, and this
should be detectable as a rise in $\sfrff$ at the typical density of
protoclusters.

Simulations based on these premises allow us to compare these models
to the observational data. Before discussing individual simulations,
we note that we cannot analyze the simulation data as thoroughly as we
have analyzed the observational data. Ideally, one would take a
snapshot of a simulation at various points in time, compute the mass
$M(>\rho)$ that is denser than $\rho$, make a plot of
$M(>\rho)/t_{\rm ff}(\rho)$, and compare that to the instantaneous
star formation rate in the simulation. This is effectively what we
have done with the observational data. However, simulation papers
generally do not report $M(>\rho)$. They only give a single density,
usually the initial (uniform) density, and report a total mass of
stars formed over the course of the entire simulation. Thus we perform
our calculations using $M(>\rho)$ for the initial density at the
initial time, and compare the mean star formation rate over the entire
simulation. This should give a rough average value that can stand in
for a more detailed calculation of $M(>\rho)/t_{\rm ff}(\rho)$ as a
function of density and time. However, we encourage authors of future
simulations of GMC evolution or star cluster formation to compute
$M(>\rho)/t_{\rm ff}(\rho)$ for at least a few time slices to
facilitate comparison to observations.

\citet{clark05} estimate from their
simulations that unbound GMCs would convert $5-10\%$ of their mass
into stars in $2-3$ free-fall times, giving $\sfrff\sim 0.03$. In
contrast, \citet{bonnell03} simulate a marginally-bound $1000$ $\msun$
clump with an initial radius of $0.5$ pc, giving $n_H=5.5\times 10^4$
cm$^{-3}$, roughly the properties of an HCN clump. There is no
feedback, and the turbulence decays freely. They find that,
after $2.6$ initial free-fall times, $58\%$ of the mass has gone into
stars, giving $\sfrff=0.22$. This is probably an underestimate of the
true value of $\sfrff$ in this scenario, since the simulation starts
with a uniform density and thus star formation does not really begin
for roughly 1 free-fall time. If we instead measured from the time
when the first star formed to the time when the last one formed, we
would find an even larger value of $\sfrff$.

This is consistent with other
simulations in which the turbulence is weak. For example,
\citet{klessen00a} model formation of a star cluster by simulating a
periodic box in which the gravitational potential energy greatly
exceeds the kinetic energy, and find that their simulation converts
$60\%$ of the initial mass into stars in 1.8 free-fall times, giving
$\sfrff=0.33$. The simulation is scale-free, but Klessen \& Burkert
suggest that the model should be reasonable for a region with a
typical density of $n_H\sim 10^5$ cm$^{-3}$. \citet{vazquezsemadeni03}
perform a parameter study of regions with turbulent driving of various
strengths and find that when the turbulent driving is weak so that the
system is unstable to global collapse, $\sfrff \sim 0.3$ is a typical
value. (As we discuss below in \S~\ref{virturb},
\citealt{vazquezsemadeni03} find that if the turbulent driving is
strong, the simulations produce lower values of $\sfrff$
that agree much better with the observations.)

We place points from the simulations of \citet{klessen00a},
\citet{bonnell03}, and \citet{clark05}, and from one of the weakly-driven
turbulence simulations of \citet{vazquezsemadeni03}, on
Figure \ref{sfrffn}. This indicates a clear observational
problem for these models. They produce roughly the correct star
formation rate at the GMC scale, but are inconsistent with the data on
the IRDC and HCN scale at the order of magnitude level. They are
consistent with the CS data only if the value of $\sfrff$ in CS clumps
is near its upper limit. We conclude
that models where star formation occurs in unbound GMCs but freely
collapsing dense clumps, or where the turbulence in those clumps is
driven but at very sub-virial levels, are ruled out by the data. Star
formation in dense clumps cannot occur through global collapse.

We emphasize again that one cannot avoid this problem by
hypothesizing a feedback process that rapidly destroys clumps once
they have turned a relatively small fraction of their mass into
stars, but retaining the picture of star formation occuring in
clumps that are undergoing global collapse. Even if feedback destroyed
collapsing clumps, while they existed they would still be forming
stars at a rate much larger than $\sfrff$ of a few percent, which is
inconsistent with the data. Only if destruction by feedback prevents
global collapse from starting can it produce the observed value of
$\sfrff$. In this case, though, estimates that cluster-forming clumps
turn $\sim 30\%$ of their gas into stars \citep{kroupa01b, lada03}
require that global collapse be held off for much longer than a
free-fall time. We discuss the implications of this in more detail in
\S~\ref{clustertime}.

One final caveat worth mentioning in our interpretation of
these simulations is that once a significant fraction of the particles
in an SPH simulation have been accreted by sink particles, the
effective resolution of that code, and its ability to follow the
hydrodynamics, may be significantly degraded. It is unclear how large
this effect is, and whether it would tend to increase or
decrease $\sfrff$. However, since the effect is only a problem late in
the simulations, it seems unlikely that it could shift $\sfrff$ by
more than a few percent.

\subsection{Galactic-Scale Gravitational Instability Models}

The criticism of unbound GMC models applies in a much weaker form to
models in which the star formation rate is determined by large-scale
global gravitational instability in a galactic disk
\citep[e.g.][]{li05a,li05b,li06,tasker06}. These models postulate that
the star formation rate in galaxies is determined by the rate at which
gravitational instability in the galactic disk drives gas to high
densities, and that the strength of this gravitational instability
determines the star formation rate. This is probably at least partly
true, in that the star formation rate is set partly by the supply of
molecular gas, which is in turn controlled by gravitational
instability. However, roughly one third of the mass of the Milky Way's
ISM is in molecular clouds. Clearly the rate-limiting step in star
formation is not at the stage of converting atomic ISM to molecular
gas, but somewhere at higher density. For this reason, the
gravitational instability models provide only a partial description
of the star formation process.

In practice, this limitation is reflected in the star formation
prescriptions that simulations of galactic-scale gravitational
instability adopt. These simulations cannot resolve structures of
molecular cloud size or smaller. Instead, when such structures form
the simulations use a recipe to determine the amount of gas that
turns into stars. These prescriptions only reproduce the correct star
formation rate if they correspond to rates $\sfrff \ll 1$ in the dense
gas.

For example, \citet{tasker06} assume for their highest resolution
simulations that in cells with densities above $10^3$ cm$^{-3}$, 50\%
of the gas is converted into stars per free-fall time of that
gas. This corresponds to $\sfrff = 0.5$. However, examining their
Figure 8 shows that this leads to a star formation rate that is an
order of magnitude larger than the observed \citet{kennicutt98a} law
without feedback, and a half an order of magnitude too large if they
include feedback. They also find that even reducing their value of
$\sfrff$ by a factor of 10 is not by itself sufficient to reproduce
the Kennicutt law. This is not surprising based on our results that
$\sfrff$ is at most a few percent in gas at densities of $10^3$
cm$^{-3}$. Reproducing the observed star formation law would require
a value for $\sfrff$ of this order.

In the simulations of \citet{li05a,li05b,li06}, the authors are able
to reproduce the \citet{kennicutt98a} law by adopting a prescription
for star formation whereby collapsed mass is assumed to form stars
with a 30\% ``efficiency'', in the sense that they count $30\%$
of the gas that is accreted onto sink particles in the simulation as
stars. The remaining 70\% of the mass stays in the sink particles, but
is assumed not to form stars. There is no simple way to
convert this recipe into a value of $\sfrff$, since it specifies
that the star formation rate in any particle is 30\% of the particle's
current accretion rate, and we do not have available the full time
history of accretion for each particle. Even if we did have this
information, interpretation would be diffuclt because the gas assumed
not to form stars remains locked up in sink particles indefinitely,
rather than being ejected back into the ISM by feedback as happens in
real star clusters. This greatly reduces the supply of gas available
to form stars. Nonetheless, the fact that these simulations need
to adopt a subgrid model that significantly reduces the collapsed mass
counted as stars in order to reproduce the \citeauthor{kennicutt98a}
law is consistent with our point that one can only explain the
observed rates of star formation if there is some mechanism that
inhibits star formation even in the densest gas that the simulations
can resolve.

The reason that these simulations reproduce the observed star
formation rate only if they have $\sfrff \ll 1$ is clear in light of
Figure \ref{sfrffn}: if the gas in IRDCs or HCN gas were collapsing on
its free-fall time, the star formation rate in the Milky Way would
exceed the observed value by two orders of magnitude. To obtain the
correct overall star formation rate, one must adopt a star formation
rate $\sfrff\ll 1$ even in the densest gas that the simulations are
currently capable of resolving. While this is an entirely reasonable
approach in simulations, a theoretical understanding of the star
formation rate requires an explanation why $\sfrff\ll 1$ even in gas
that is already collapsed to densities above the critical density of
HCN(1-0), five orders of magnitude higher than the mean in the
Galactic ISM. This point does not mean that global gravitational
instability is unimportant in regulating star formation, simply that
it cannot be the sole agent. Thus, models such as these are
complementary to models of $\sfrff$ in dense gas such as those of
\citet{krumholz05c}. 

\subsection{Magnetic Regulation Models}

Another possible explanation for low star formation rates is strong
magnetic fields \citep[e.g.][]{shu87, mckee89, tassis04,
nakamura05a}. If star-forming clouds are magnetically subcritical,
which is controversial on both observational and theoretical grounds
\citep{mckee93,crutcher99,bourke01,padoan04b,heiles05}, then they
cannot collapse before
ambipolar diffusion allows the magnetic field to slip out of the gas,
and star formation proceeds on the ambipolar diffusion time scale
instead of the free-fall time scale. For a uniform gas, the
time scale required for ambipolar diffusion to decouple the gas and
the field is \citep{shu92}
\begin{equation}
\label{tadest}
\tad \sim \frac{L^2}{v_A^2} \gamma C \rho^{1/2},
\end{equation}
where $L$ and $\rho$ are the characteristic size scale and density of the
cloud, $v_A$ is the Alfv\'en speed, $\gamma\approx 3.5\times 10^{13}$
cm$^3$ g$^{-1}$ s$^{-1}$ is the ion-neutral drag coefficient, and
$C=3\times 10^{-16}$ cm$^{-3/2}$ g$^{1/2}$ is the cosmic ray
ionization constant. (In high column density environments like IRDCs,
interstellar UV photons cannot penetrate and ionization is dominated
by cosmic rays.) If we assume that this describes the star formation
timescale, then the star formation rate should obey roughly
\citep{ciolek93}
\begin{equation}
\label{sfrffmag}
\sfrff\sim \frac{\ecore \tff}{\tad},
\end{equation}
where $\ecore\approx 0.5$ is the fraction of the mass in a prestellar
core that reaches the star rather than being ejected by outflows
\citep{matzner00}.

To determine what this implies for $\sfrff$, we
need to know how the magnetic field behaves in objects of varying
densities. While there are very few direct observations of magnetic
fields for extremely dense objects of the sort we are considering,
observations do show that magnetic fields obey the correlation $B =
(8\pi)^{1/2} \sigma \rho^{1/2} \mu^{-1} c_1^{1/2}$, where $\sigma$ is
the velocity dispersion of the region, $\mu$ is the ratio of the
object's mass-to-flux ratio to the critical value $(2\pi
G^{1/2})^{-1}$, $c_1$ is a constant of order unity the depends on
the cloud's internal density distribution, and $\mu^{-1}
c_1^{1/2}\approx 0.8$ \citep{basu00}. This correlation is what one
would expect if, in one direction, cloud self-gravity were balanced by
magnetic plus turbulent pressure. If we use this correlation in
(\ref{sfrffmag}), and re-write the relation in terms of the virial
parameter $\avir\equiv 5 \sigma^2 L/(G M)$, we find that all the
dependence on dimensional quantities drops out and we are left with
\begin{equation}
\sfrff \sim 0.01 \avir.
\end{equation}

The uncertainty of this calculation is probably more than an order of
magnitude,
so we should not pay particular attention to the coefficient, and we
will not attempt to place points for magnetic regulation models on
Figure \ref{sfrffn}. Major contributors to the uncertainty are
ambiguities in the definition of the length and mass scales $L$ and
$M$, the lack of three-dimensional numerical simulations to determine
how well equation (\ref{sfrffmag}) holds on scales larger than a core
collapsing to form a single star system, and the possibility that the
cosmic ray ionization rate may vary substantially between galaxies.
Perhaps most significantly, equation (\ref{tadest}) assumes that
ambipolar diffusion in a turbulent medium is not substantially faster
than in a quiescent one, an assumption that may well fail
\citep[e.g.][]{heitsch04, nakamura05a}. If turbulence shortens the
ambipolar diffusion time scale so that $t_{\rm AD}\sim t_{\rm ff}$, as
some simulations suggest it might, then magnetic regulation models
would produce a value of $\sfrff$ that is too high. Whether this
occurs or not is a question that will have to be addressed through
three-dimensional simulations with self-gravity and non-ideal MHD.

Given these uncertainties, we cannot really say whether the rate of
star formation in magnetically subcritical clouds subject to ambipolar
diffusion is quantitatively consistent with the observations. However,
the lack of dependence of $\sfrff$ on any properties but the virial
parameter implies that we expect the magnetic $\sfrff$ to be roughly
the same in all virialized objects, a prediction we can compare to
observations. For our observed objects, we know that GMCs with masses
$\gtsim 10^4$ $\msun$, HCN clumps, and CS clumps are roughly virialized
\citep{plume97, heyer01,wu05}, so $\avir\approx 1$ for them. For IRDCs we
lack kinematic information from optically thin molecular emission, and
therefore we cannot directly determine the velocity dispersion and the
virial parameter. If we assume that these objects are virialized, then
the magnetic regulation model is broadly consistent with observations
that $\sfrff$ is roughly constant. We cannot make a stronger statement
than this because the magnetic regulation model cannot currently make
more specific predictions. Note that it is not critical for this
purpose that the objects in question truly be in exact virial
equilibrium, since they almost certainly are not. We are using the
virial parameter only as a way of parameterizing the strength of the
turbulence relative to the strength of gravity, so all that we require
is that the objects have virial parameters $\avir\sim 1$, i.e. that
they not be either completely collapsing or completely
non-self-gravitating.

\subsection{Virialized Turbulence Models}
\label{virturb}

A fourth idea to explain the low star formation rate is that
turbulence inhibits collapse. This idea has a long history, and we
refer readers to \citet{maclow04} for a thorough review.
Recently, \citet{krumholz05c} synthesized simulations and analysis of
turbulence-inhibited star formation by \citet{klessen00b} and
\citet{vazquezsemadeni03} to derive an estimate that turbulence
produces a star formation rate $\sfrff\approx 0.014
(\avir/1.3)^{-0.68} (\calm/100)^{-0.32}$, where $\avir$ and $\calm$
are the virial parameter and Mach number of a star-forming gas
cloud. The estimate is based on a derivation of the fraction of mass
that is unstable to collapse in a medium that has the density and
velocity structure common to supersonic isothermal turbulence, and is
calibrated against simulations. Its uncertainty is probably a factor
of a few, stemming from uncertainty in the effects of magnetic fields
and from the uncertain approximation $\ecore=0.5$. For GMCs, HCN
clumps, and CS clumps, typical values are $\avir \sim 1-2$ and
$\calm\sim 20-40$
\citep{solomon87,mckee99a,plume97,shirley03,wu05}. As with the
magentic models, exact virial equilibrium is not required, simply that
the kinetic and potential energies be comparable. We plot the
predicted value of $\sfrff$ for this range of parameters in Figure
\ref{sfrffn}. Given the uncertainties in both the observations and the
theoretical calculation, there is reasonable agreement.

Simulations of star formation in turbulent gas for which the turbulent
energy is roughly equal to the gravitational potential energy, either
because it is driven artificially or because the simulations include
feedback from protostellar outflows, also give values of $\sfrff$ that
are broadly consistent with the data. We show three examples in Figure
\ref{sfrffn}: 
\citet{li04} simulate a periodic box in which the kinetic and
turbulent energies are approximately equal and the turbulence is
driven to keep the level of turbulence roughly constant. They find a
star formation rate of $\sfrff=0.079$ at a resolution of $512^3$
cells. (Since the simulation uses periodic boundary conditions, the
true gravitational potential energy is not well-defined, so we cannot
state that the virialization is more than approximate.) The
simulations are dimensionless, but they suggest scaling to a typical
density of $2\times 10^5$ cm$^{-3}$. Similarly, in scale-free
simulations with strong turbulent driving, \citet{vazquezsemadeni03}
and \citet{vazquezsemadeni05b} find $\sfrff\sim 0.01$. \citet{li06b}
find
$\sfrff=0.067$ in a simulation with a central density of $5\times
10^4$ cm$^{-3}$ that includes protostellar outflows
from forming stars, which keep the kinetic and gravitational potential
energies approximately equal. A number of other simulations in which
the turbulence is roughly in virial balance with the gravitational
energy produce star formation rates in reasonable agreement with
observations over the period in the simulations for which that balance
is maintained. If the turbulence is sufficiently strong, simulations
can achieve $\sfrff\sim 0.01$ \citep[e.g.][]{vazquezsemadeni03,
tilley04, vazquezsemadeni05b}. 

We note that $\sfrff$ by itself does not appear to provide a way of
distinguishing the length scale at which turbulence is driven, since
simulations in which turbulence is driven primarily at large scales
and ones in which it is driven primarily at small scales both seem
capable of producing low values of $\sfrff$. Some authors
\citep[e.g.][]{vazquezsemadeni03} emphasize the length scale of the
driving as critical to determining whether the turbulence can keep
$\sfrff$ low, and \citet{krumholz05c} show that the
characteristic distinguishing driving that gives $\sfrff\sim 1$ from
driving that produces $\sfrff \ll 1$ is how the Jeans length in the
gas compares to the sonic length of the turbulence, defined as the
length scale for which the turbulent velocity dispersion equals the
sound speed. Other diagnostics, such as the IMF and clustering
statistics \citep[e.g.][]{klessen01, schmeja04}, the
linewidth-size relation of the gas \citep[e.g.][]{ossenkopf02}, and
the line luminosity of molecular clouds \citep{heyer04} suggest
that turbulence in molecular clouds is primarily driven on large
scales. In this case \citeauthor{krumholz05c} show that the condition
distinguishing cases with $\sfrff \sim 1$ from those with $\sfrff \ll
1$ reduces to the statement that clouds with $\avir\gtsim 1$ have low
values of $\sfrff$, while those with $\avir \ll 1$ have $\sfrff
\sim 1$. Thus, the combination of observations of $\sfrff$ and those
showing that most turbulent energy is on large scales imply that
turbulence can only reproduce the observed values of $\sfrff$ if it
is driven strongly enough to produce $\avir\sim 1$.

\subsection{The Formation Timescale of Clusters}
\label{clustertime}

Another important implication of this work is for the formation
timescale of star clusters. The star formation efficiency of clusters,
defined as the fraction of the initial gas mass that forms stars over
the lifetime of the gas clump, is thought to be $\sim 20\%-50\%$
\citep{kroupa01b,lada03}. Since the observational data show that
$\sfrff$ is at most a few percent, this implies that the cluster
formation process must take at least $\sim 10\tff$ ($\sim 5$
crossing times).

The recent observation by \citet{wu05} that individual HCN clumps in
the Galaxy with IR luminosities $L_{\rm IR} > 10^{4.5}$ $\lsun$ lie on
the same $L_{\rm IR}-L_{\rm HCN}$ correlation as entire galaxies also
provides indirect evidence that the cluster formation time scale is
comparatively long. Wu et al. suggest that luminous HCN clumps fall on
the galactic IR-HCN correlation because they contain enough stars to
sample the IMF fully. Our results showing a large scatter in the light
output for small clusters support this conjecture. However, our
results on the age-dependence of the luminosity suggest that a mass
large enough to sample the IMF is not sufficient by itself to place a
cluster on the extragalactic IR-HCN correlation. Since star formation
in the galaxies surveyed by \citet{gao04b,gao04a} has been ongoing for
$\gtsim 3$ Myr, their IR luminosities trace the
star formation rate, and the $L_{\rm IR}-L_{\rm HCN}$ correlation is
therefore a statement of the star formation rate per unit mass in HCN
gas. If the Galactic HCN clumps observed by \citet{wu05} typically
survived for only $\sim 1$ free-fall time, $\sim 0.2$ Myr, then as
Figure \ref{lt} shows their luminosity per unit star formation rate
would be quite different than that of an older galactic population,
and they would not lie on the galactic correlation. One would only
expect Galactic HCN clumps to follow the extragalactic $L_{\rm
IR}-L_{\rm HCN}$ correlation if they are massive enough to sample the
IMF well \textit{and} if they are $\gtsim 1$ Myr old, which means that
they must have existed for at least $\sim 5\tff$, and probably closer
to $10\tff$.

These two lines of evidence provide additional support for
the argument that star clusters form in near-equilibrium gas clumps
presented by \citet{tan06a}. Tan et al. discuss the implications of
this finding in more detail, but we note here two of the most
significant. First, formation times of $\sim 10 \tff$ or longer appear
inconsistent with models in which turbulence in protocluster gas is
freely decaying. Simulations show that if turbulence is allowed to
decay freely \citep[e.g.][]{bonnell03} or never contained energy
comparable to the gravitational potential energy
\citep[e.g.][]{vazquezsemadeni03}, the star formation process
generally ends in only about $2\tff$, with $\sim 50\%$ of
the mass in stars. Only if the turbulence is continually driven can
there be enough gas left up to $10\tff$ for
the star formation process to continue
\citep[e.g.][]{vazquezsemadeni03,li04,li06b}. The long cluster
formation time scales suggested by the observations therefore imply
that either the simulations of 
decaying turbulence are incorrect or (more likely) the turbulence is
continually driven. This is consistent with on observations by
\citet{williams03} and \citet{quillen05}, and with simulations of
protostellar feedback by \citet{li06b}. We note again that the
observations of $\sfrff$ alone put no constraints on how or at what
scale the turbulence is driven. It can be from internal sources of
feedback on small scales, or from a turbulence cascade in which most
of the power is on the size scale of an entire GMC.

A second implication is for the
mechanism responsible for the stellar initial mass
function. \citet{krumholz05e} have recently emphasized that the
competitive accretion mechanism \citep[and references
therein]{bonnell06a} can only operate in the context of a strongly
sub-virial gas clump that is undergoing global collapse and converting
order unity of its gas mass into stars in a free-fall
time. \citet{bonnell06c} concur that the question of whether
competitive accretion occurs or not depends on the extent to which gas
and stellar velocities in a forming star cluster remain tightly
coupled for the majority of the formation process, which in turn
depends on whether feedback is effective at driving turbulence, which
would change gas velocities but not stellar velocities. If turbulent
driving keeps clusters virialized for many free-fall times and limits
$\sfrff$ to $\ltsim 0.1$, then competitive accretion cannot occur
within those clusters. The observations we discuss here provide
evidence for low values of $\sfrff$ and long cluster formation times,
and therefore provide strong evidence against the possiblity that
competitive accretion determines the IMF.

\section{Conclusions}
\label{conclusions}

We present observational evidence for two
surprising conclusions, one shown quite strongly and the other more
tentatively. First, and very clearly, \textit{star formation in dense
gas is slow.} The time required to convert all the gas into stars, the
depletion time, is longer than the free-fall time by at least and
order of magnitude, and probably closer to two orders of
magnitude. There is no evidence that gas at densities up to
$n_H \sim 10^5$ cm$^{-3}$ resides primarily in objects that are
undergoing collapse and rapid star formation. Second, and more
tentatively, \textit{the ratio of free-fall time to the depletion time
is independent of the characteristic
density of the star-forming object in question}. The data at present
are uncertain, and are insensitive to changes of less than roughly an
order of magnitude. Nonetheless, it is interesting that this ratio
varies by no more than an order of magnitude from the typical density
of GMCs to the typical density of HCN(1-0)-emitting gas, a range of
nearly 3 decades in density.

These observations are a strong constraint for theories of star
formation, and seem difficult to explain in the context of models in
which there is a transition from slow, unbound star formation to
rapid, bound star formation somewhere between the GMC scale and the
protocluster scale. Slow star formation in cluster-forming gas also
implies that clusters require many free-fall times to assemble, as
recently argued by \citet{tan06a} on other grounds. Models in which
star formation takes place in virialized objects and is inhibited by
strong magnetic fields are qualitatively consistent with the data, and
models in which star formation is inhibited by turbulence are both
qualitatively and quantitatively consistent with observations.

In the future it would be extremely useful to improve the data on which 
Figure \ref{sfrffn} is based. One way to do this would be to perform
detailed studies of other young clusters and obtain data comparable to
that for the ONC. This would provide a method of estimating $\sfrff$
that is independent of luminosity conversions and does not suffer from
concerns about the completeness of galactic surveys. Another
improvement in the data could come from
an unbiased survey of CS(5-4) emission in the Milky Way or in another
galaxy, which would allow us to replace the upper limit on
$\sfrff$ we derive here with an actual estimate. This would be
particularly valuable because the upper limit on $\sfrff$ from CS(5-4)
is well above the estimate from HCN(1-0), even though the densities
are not very different. Unbiased CS observations could likely bring
down this point.

Another improvement would be extend the data to higher
densities. To accomplish this will require observations
either of external galaxies or relatively complete surveys of the
Milky Way in molecular transitions that trace densities $\gtsim 10^6$
cm$^{-3}$. Determining masses from the luminosities in these
transitions will probably require high resolution follow-up
observations of Galactic sources in optically thin isotopomers so that
the luminosity may be correlated against a virial mass. While this
is a significant observational challenge, such surveys might make it
possible to identify a scale at which star formation
transitions from slow to fast, a crucial datum in understanding the
star formation process.

\acknowledgements We thank N. Evans, C. McKee, and J. Wu for helpful
discussions, and the referee, Ralf Klessen, for comments that improved
the quality of the paper.
Support for this work was provided by NASA through Hubble
Fellowship grant HSF-HF-01186 awarded by the Space Telescope Science
Institute, which is operated by the Association of Universities for
Research in Astronomy, Inc., for NASA, under contract NAS 5-26555
(MRK).

%\bibliographystyle{apj}
%\bibliography{refs}

%--------------------------------------------------------------------------
\end{document}